\documentclass[reqno]{amsart}
\def\preprint{}

\title[Fixed-point elimination in the \Ipc]{Fixed-point elimination in the \\Intuitionistic Propositional
  Calculus} 

\author[Ghilardi]{Silvio Ghilardi}
\address{Silvio Ghilardi\\
 Dipartimento di Matematica, Universit\`a degli Studi di
  Milano}
\email{silvio.ghilardi@unimi.it}

\author[Gouveia]{Maria Jo\~{a}o Gouveia} 
\address{Maria Jo\~{a}o Gouveia \\
 CEMAT-CI\^ENCIAS, 
 Universidade de Lisboa, 
1749-016, Lisboa, Portugal}
\email{mjgouveia@fc.ul.pt}

\author[Santocanale]{Luigi Santocanale}
\address{Luigi Santocanale\\
LIF, CNRS UMR 7279, Aix-Marseille Universit\'e}
\email{luigi.santocanale@lif.univ-mrs.fr}

\newcommand{\cal}[1]{\mathcal{#1}}
\input macros.tex

\renewenvironment{proof}{\par\noindent\emph{Proof.}~}{\par\noindent}

\begin{document}
\maketitle
\begin{abstract}
  It is a consequence of existing literature that \lgfp{s} of monotone
polynomials on \Ha{s}---that is, the algebraic models of the
\IPC---always exist, even when these algebras are not complete as
lattices.  The reason is that these \efp{s} are definable by formulas
of the \Ipc. Consequently, the $\mu$-calculus based on intuitionistic
logic is trivial, every $\mu$-formula being equivalent to a
fixed-point free formula. We give in this paper an axiomatization of
\lgfp{s} of formulas, and an algorithm to compute a \fpf formula
equivalent to a given $\mu$-formula. The axiomatization of the \gfp is
simple. The axiomatization of the \lfp is more complex, in particular
every monotone formula converges to its \lfp by Kleene's iteration in
a finite number of steps, but there is no uniform upper bound on the
number of iterations. We extract, out of the algorithm, upper bounds
for such $n$, depending on the size of the formula.  For some
formulas, we show that these upper bounds are polynomial and optimal.

\end{abstract}

\section{Introduction}

In \cite{Ruitenburg84} the author proved that, for each formula
$\phi(x)$ of the \IPC, there exists a number $n \geq 0$ such that
$\phi^{n}(x)$---the formula obtained from $\phi$ by iterating $n$
times substitution of $\phi$ for the variable $x$---and
$\phi^{n+2}(x)$ are equivalent in \IL. This result has, as an
immediate corollary, that a syntactically monotone formula $\phi(x)$
converges both to its \lfp and to its \gfp in at most $n$ steps.
Using a modern notation based on $\mu$-calculi \cite{AN01}, we have
$\mu_{x}.\phi(x) = \phi^{n}(\bot)$ and $\nu_{x}.\phi(x) =
\phi^{n}(\top)$.  These identities also show that a $\mu$-calculus
based on \IL is trivial, every $\mu$-formula being equivalent to a
\fpf formula.

Ruitenberg's work \cite{Ruitenburg84} leaves open how to extract or
estimate the least number $\rho(\phi)$ such that $\phi^{\rho(\phi)}(x)
= \phi^{\rho(\phi)+2}(x)$. Yet, our motivations stem from the theory
of \efp{s} and $\mu$-calculi \cite{AN01}. In principle, being able to
compute or bound Ruitenberg's number $\rho(\phi)$ might end up in an
over-approximation of the closure ordinal of $\phi$---the least $k$
such that $\mu_{x}.\phi(x) = \phi^{k}(\bot)$. For the analogous
problem with the \gfp, we shall see that the least number $k$ such
that $\nu_{x}.\phi(x) = \phi^{k}(\top)$ is bounded by $1$, while
$\rho(\phi)$ might be arbitrarily large.

Later in \cite{Mardaev1993}, the author gave an independent proof
that \lfp{s} of monotone formulas are definable within \IL.  His proof
relies on semantics methods and on the coding of \IL into \GRZL; the
proof was further refined in \cite{Mardaev1994} to encompass the
standard coding of \IL into its modal companion, the logic \Sfour.
Curiously, no mention of \gfp{s} appears in these works.

Another relevant source for this paper stem from the discovery that
\Ipc has uniform interpolants \cite{Pitts92}, often named
\emph{bisimulation quantifiers}. Together with the deduction property
of \Ipc, they give the category of (finitely generated) \Ha{s}---that
is, the algebraic models of the \IPC---a rather strong structure,
axiomatized and studied in \cite{GhilardiZawadowski2011,GhilardiZawadowski97}. It is possible
to argue that in every category with similar properties the \efp{s} of
monotone formulas are definable. This is possible by using quantified
formulas analogous to the one used in \cite[\S 3]{DAgostinoHollenberg2000}
to argue that \PDL lacks the uniform interpolation
property. 
In this paper we exploit this idea and the existential bisimulation
quantifiers to characterize \gfp{s} in the \IPC.

\bigskip

A $\mu$-calculus is
a prototypical kind of computational logic, obtained from a base logic
or algebraic system by addition of distinct forms of iteration so to
increase expressivity.
This paper is part of a line of research whose goal is to understand,
under a unified perspective, why alternation-depth hierarchies of
$\mu$-calculi are degenerate or trivial.
A $\mu$-calculus adds to an underlying logical-algebraic system formal
lgfp{s} of \fterms whose semantic monotonicity can be witnessed at
the syntactic level. When addition of \efp{s} is iterated, \fterms
with nested \efp{s} are generated. The alternation-depth hierarchy
\cite[\S 2.6]{AN01} of a $\mu$-calculus measures the complexity of a
\fterm as a function of the nesting of the different types of \fp{s},
with respect to a fixed class of models.
It is well known that \fp{s} that are unguarded can be eliminated in
the propositional modal $\mu$-calculus \cite{Kozen83}. We can rephrase
this fact by saying that the alternation-depth hierarchy of the
$\mu$-calculus over distributive lattices is
trivial, every $\mu$-term being equivalent to a \fpf term. A goal of
\cite{FrittellaSantocanaleRAMICS} was to understand closely this
result and to generalize it. We were able to exhibit equational
classes of lattices ${\cal D}_{n}$---with ${\cal D}_{0}$ the class
of distributive lattices---where the \efp{s} can be uniformly computed
by iterating a \fterm $n+1$ times from the bottom/top of the lattice;
moreover, we showed that these uniform upper bounds are optimal. The
reasons for the degeneracy of the hierarchy can be ultimately found in
the structural theory of lattices.

As we show in this paper, the situation is quite different when the
base for the $\mu$-calculus is \IL, with its standard models the
\Ha{s}. 
Several ingredients contribute to the existence of a closure ordinal
of each formula and to its finiteness. Among them, \emph{strongness}
of the monotone polynomials on \Ha{s}. This means that a monotone
polynomial $\f : H \rto H$ over a \Ha $H$ can be considered as a
functor enriched over $H$, when $H$ is consider as a closed category
\cite{Kelly82}.  
For some polynomials, existence and finiteness of the closure ordinal
is a consequence of being \emph{inflating} (or expanding) and, on the
syntactic level, to a restriction to the use of conjunction that
determines a notion of disjunctive formula.
As far as the \gfp is concerned, monotone formulas uniformly converge
to it after one step. A key ingredient of the algorithm we present is
creation of \lfp{s} via the \emph{Rolling} equation
(cf. Lemma~\ref{lemma:rolling}), a fact already used in
\cite{DAgostinoLenzi2010}. For \IL and \Ha{s}, where \fterms can be
semantically antitone (i.e. contravariant), existing \gfp{s} create
\lfp{s}.
The most striking difference with the case of distributive lattices
(and with the case of the varieties ${\cal D}_{n}$) is the absence of
a finite uniform upper bound on the closure ordinals,
the rate of convergence to the
\lfp crucially depending on the shape 
of the formula.

\medskip

As emphasized in \cite{LehtinenQuickert15} for the propositional modal
$\mu$-calculus, once a formula is known to be equivalent to some other
formula of smaller complexity, we should also be able to effectively
compute this second formula. Thus, the fact that the alternation
hierarchy is trivial for $\mu$-calculi based on the \Ipc
should not be the end of the story. The main contribution of this
paper is to achieve an effective transformation of an intuitionisitc
$\mu$-formula into an equivalent \fpf intuitionisitc formula.
The size of the formula might exhibit an exponential grow during this
transformation. Yet, this is mainly due, as usual, to the 
need of precompiling a formula into an equivalent one in some kind of
conjunctive normal form.
We might use sharing in substitutions---or introduce the appropriate
formalism for approximants to \lfp{s}---so that, if we are given an
already precompiled formula, then its \lfp w.r.t. the variable $x$ is
only polynomially bigger than the original formula.
For these formulas, we instantiate this claim by explicitly giving a
way of computing $f(\phi)$ such that $\mu_{x}.\phi(x) =
\phi^{f(\phi)}(\bot)$, so that $f(\phi)$ is an upper bound to the
closure ordinal of $\phi$.
In some cases we are able to show that $f(\phi)$ is optimal, by
exhibiting some formula $\phi(x)$ such that $\phi^{f(\phi) -1}(\bot)
<\mu_{x}.\phi(x) $.

\bigskip

The paper is structured as follows. We recall in
Section~\ref{sec:notation} some elementary facts from \fp theory. In
Section~\ref{sec:intmucalculus} we recall the \IPC and introduce the
\muIPC.  In Section~\ref{sec:strongfunctions} we argue that monotone
polynomials are strong and exhibit the interactions between \lfp{s}
and strong functions. In Section~\ref{sec:quantifiers} we use the
existential bisimulation quantifier to argue that monotone polynomials
converge to their \gfp in one step. Section~\ref{sec:procedure} is the
core of our paper, where we show ho to eliminate a \lfp from a
formula. Together with the result in the previous Section, this leads
to a procedure to eliminate off the \fp{s} from a \muIpc formula.
Finally, in Section~\ref{sec:upperbounds}, we show how upper bounds to
closure ordinals can be extracted from the procedure elimination of
the \lfp{s}. In 
Section~\ref{sec:conclusions} we present our final remarks.

\section{Notation and elementary concepts}
\label{sec:notation}

Let $P$ and $Q$ be posets. A function $f : P \rto Q$ is
\emph{monotone} if $x \leq y$ implies $f(x) \leq f(y)$, for each
$x,y \in P$.  If $f : P \rto P$ is a monotone endofunction, then
$x \in P$ is a \emph{\pfp} of $f$ if $f(x) \leq x$; we denote by
$\Pref_{f}$ the set of prefixed points of $f$. 
Whenever $\Pref_{f}$ has a least element, we denote it by $\mu.f$.
Therefore, $\mu.f$ denotes the \emph{\lpfp} of $f$, whenever it
  exists.  If $\mu.f$ exists, then it is a \fp of $f$, necessarily
the least one.  
The notions of \lpfp and of \lfp
coincide on complete lattices
or when the \lfp is computed by
iterating from the bottom of a lattice; for our purposes they are
interchangeable, so we shall abuse of language and refer to $\mu.f$ as
the \lfp of $f$.
Dually (and abusing again of language), the \emph{\gfp} of $f$ shall
be denoted by $\nu.f$.

\smallskip

Let us mention few 
elementary facts from \fp theory.
\begin{lemma}
  \label{lemma:rolling}
  Let $P, Q$ be posets, $f : P \rto Q$ and $g : Q \rto P$ be monotone
  functions. If $\mu.(g \circ f)$ exists, then $\mu.(f \circ g)$
  exists as well and is equal to $f(\mu.(g \circ f))$. 
\end{lemma}
As we do not work in complete lattices (so we are not ensured that
\lfp{s} exist) we express the above statement via the equality
\begin{align}
  \label{eq:rolling}
  \tag{\texttt{Roll}}
  \mu.(f \circ g) & := f(\mu.(g \circ f))\,,
\end{align}
where the colon emphasizes existence: if the \lfp in the expression on
the right \emph{exists}, then this expression is the \lfp of
$f\circ g$. Analogous notations will be used later. We endow the
  product of two posets $P$ and $Q$ with the coordinatewise ordering.
  Therefore a function $f : P \times Q \rto R$ is monotone if, as a
  function of two variables, it is monotone in each variable. To deal
with \lfp{s} of functions of many variables, we use the standard
notation: for example, if $f : P\times P\rto P$ is the monotone
function $f(x,y)$, then, for a fixed $p \in P$, $\mu_{x}.f(x,p)$
denotes the \lfp of $f(x,p)$. Let us recall that the correspondence
$p \mapsto \mu_{x}.f(x,p)$---noted $\mu_{x}.f(x,y)$---is again
monotone.
\begin{lemma}
  \label{lemma:diag}
  If $P$ is a poset and $f : P\times P \rto P$ is a monotone mapping,
  then
  \begin{align}
    \label{eq:diag}
    \tag{\texttt{Diag}}
    \mu_{x}.f(x,x) & := \mu_{x}.\mu_{y}.f(x,y)\,.
  \end{align}
\end{lemma}
\begin{lemma}
  \label{lemma:bekic}
  If $P$ and $Q$ are posets and $\langle f,g\rangle : P\times Q \rto
  P\times Q$ is a monotone function, then $\mu.\langle f,g \rangle :=
  \langle \mu_{1},\mu_{2}\rangle$, where
  \begin{align}
    \label{eq:bekic}
    \tag{\texttt{\Bekic}}
    \mu_{1} & = \mu_{x}.f(x,\mu_{y}.g(x,y)) \quad \tand\quad \mu_{2} =
    \mu_{y}.g(\mu_{1},y)\,.
  \end{align} 
\end{lemma}

\section{The  \muIPC}
\label{sec:intmucalculus}

Formulas of the \IPC are generated according to the following grammar:
\begin{align}
  \label{grammar:IL}
  \phi & \production x \mid \top \mid \phi \land \phi \mid \bot \mid \phi
  \vee \phi \mid \phi \impl \phi\,, 
\end{align}
where $x$ ranges over a countable set $\Vars$ of propositional
variables.  For the \Ipc, the formulation of the consequence relation
$\cons$ (relating a set of formulas to a formula) goes back to
Gentzen's work on the system \LJ \cite{Gentzen35}. It is well known
that \IL is sound and complete w.r.t. the class of its algebraic
models, the \Ha{s}.
\begin{definition}
  A \emph{\Ha} $H$ is a bounded lattice (with least element $\bot$ and
  greatest element $\top$) equipped with a binary operation $\impl$
  such that the following equations hold in $H$:
  \begin{align}
    \notag
    x \land (x \impl y) & = x \land y\,, &
    x \land (y \impl x) & = x\,,\\
    x \impl x & = \top \,, &
    x \impl (y \land z) & = (x \impl y) \land  (x \impl z)\,.
    \label{eq:distrimpl}
  \end{align}
\end{definition}
We can define on any \Ha a partial order by saying that $x \leq y$
holds when $x \vee y = y$.  We identify formulas of the \Ipc with
terms of the theory of \Ha{s}, constructed therefore from variables
and using the signature $\langle \top,\land, \bot,\vee, \impl\rangle$.
For $\phi$ such a \fterm, $H$ a \Ha, and $v : \Vars \rto H$ a
valuation of the propositional variables in $H$, let us write
$\eval{\phi}$ for the result of evaluating the formula in $H$,
starting from the variables.  The soundness and completeness theorem
of the \Ipc over \Ha{s}---see e.g. \cite{BJ2006}---can
then be stated as follows: \emph{if $\Gamma$ is a finite set of
  \fterms and $\phi$ is a \fterm, then $\Gamma \cons \phi$ holds if
  and only if
  $\bigwedge_{\gamma \in \Gamma} \eval{\gamma} \leq \eval{\phi}$
  holds, in every \Ha $H$ and for every valuation of the propositional
  variables $v : \Vars \rto H$.}  Given this theorem, we shall often
abuse of notation and write $\leq$ in place of $\cons$, and the
equality symbol $=$ to denote logical equivalence of formulas.

\medskip

We aim at studying \efp{s} on \Ha{s}. To this end, we formalize the
\muIPC.

An occurrence of a variable $x$ is \emph{positive} in a \fterm $\phi$
if, in the syntax tree of $\phi$, the path from the root to the leaf
labeled by this variable contains an even number of nodes labeled by
subformulas $\psi_{1} \impl \psi_{2}$ immediately followed by a node
labeled by the subformula $\psi_{1}$.  If, on this path the number of
those nodes is odd, then we say that this occurrence of $x$ is
\emph{negative} in $\phi$.  A variable $x$ is positive in a formula
$\phi$ if each occurrence of $x$ is positive in $\phi$. A variable $x$
is negative in a formula $\phi$ if each occurrence of $x$ is negative
in $\phi$.
If we
add to the previous grammar~\eqref{grammar:IL} the following
productions:
\begin{align*}
  \phi \production & \mu_{x}.\phi\,,  & \phi \production & \nu_{x}.\phi\,,
\end{align*}
subject to the restriction that $x$ is positive in $\phi$, we obtain
then a grammar for the formulas of \muIpc, the \muIPC.
The semantics of these formulas is the expected one. Let $\phi$ be a
formula of \muIpc, and let $x$ be positive in $\phi$. Let us denote by
$\Vars[\phi]$ the set of variables having an occurrence in $\phi$. If
$v : \Vars[\phi] \setminus \set{x}\rto H$ is a valuation of all the
variables of $\phi$ but $x$ in a \emph{complete} \Ha, then the
function $\eval[v]{\phi}$, defined by 
\begin{align*}
  h & \mapsto \eval[v,h/x]{\phi}\,,
\end{align*}
is monotone, so $\mu_{x}.\phi$ (resp., $\nu_{x}.\phi$) is to be
evaluated over the \lfp (resp., the \gfp) of this function. A sequent
calculus for \muIpc is presented in \cite[\S 2]{Clairambault13}.

Let us say that a formula $\phi$ of $\muIpc$ is \fpf if it is a
formula of $\Ipc$, that is, it does not contain either of the symbols
$\mu,\nu$.
\begin{proposition}
  Every formula $\phi$ of $\muIpc$ is equivalent to a \fpf formula $\phi'$.
\end{proposition}
\begin{proof}
  Clearly, the statement holds if we can show that it holds whenever
  $\phi = \mu_{x}.\psi$ or $\phi = \nu_{x}.\psi$, where $\psi$ is a
  \fpf formula.
  For a natural number $n \geq 0$, let $\psi^{n}(x)$
  denote the formula obtained by substituting $x$ for $\psi$ $n$
  times. 
  Ruitenburg \cite{Ruitenburg84} proves that, for each intuitionisitic
  formula $\psi$, there exists a number $n \geq 0$ such that the
  formulas $\psi^{n}(x)$ and $\psi^{n + 2}(x)$ are equivalent.
  If $x$ is positive in $\psi$, then instantiating $x$ with $\bot$,
  leads to the equivalence $\psi^{n+1}(\bot) \equiv \psi^{n}(\bot)$,
  exhibiting $\psi^{n}(\bot)$ as the \lfp of $\psi$. Similarly,
  $\psi^{n}(\top)$ is the \gfp of $\psi$.  \qed
\end{proof}
While it is an obvious step to derive the previous Proposition from
Ruitenburg's result, there has been no attempt (as far as we know) to
compute an upper bound on $n \geq 0$ such that $\psi^{n}(x)$ and
$\psi^{n + 2}(x)$ are equivalent. Nor is such an $n$ necessarily a
tight upper bound for convergence of a formula to its least or \gfp.

\section{Strong monotone functions and \fp{s}}
\label{sec:strongfunctions}

If $H$ is a Heyting algebra and $f : H \rto H$ is a monotone
function, then we say that $f$ is \emph{strong} if
\begin{align*}
  x \land f(y) & \leq f(x \land y)\,,
  \tag*{for any $x, y \in H$.}
\end{align*}
The interplay between \fp{s} and this class of functions has already
been emphasized, mainly in the context of categorical proof-theory and
semantics of functional programming languages with inductive data
types \cite{CockettSpencer95,Clairambault13}.

\begin{lemma}
  \label{lemma:strong}
  A monotone $f : H \rto H$ is strong if and only if any of the following
  equivalent conditions holds in $H$:
  \begin{align}
    x \land f(y) & \leq f(x \land y)\,, \label{eq:strongconj} \\
    f(x \impl y) & \leq x \impl f(y)\,, \label{eq:strongimpl} \\
    x \impl y & \leq f(x) \impl f(y)\,. \label{eq:strongenriched}
  \end{align}
\end{lemma}
The proof of these equivalences is usual in categorical algebra
\cite{Kock72} and therefore it is omitted here.

\medskip

\begin{definition}
  \label{def:monotonepol}
  Let $H$ be a Heyting algebra. We say that a function $f : H \rto H$
  is \emph{monotone polynomial} if there exist a formula $\phi$ of the
  \Ipc, a variable $x$ positive in $\phi$, and a valuation $\vec{v} :
  \Vars[\phi] \setminus \set{x} \rto H$ such that, for each $h \in H$,
  we have $f(h) = \eval[\vec{v},h/x]{\phi}$.
\end{definition}

\begin{proposition}
  \label{prop:Peirce}
  Every monotone polynomial $\f$ on a \Ha is strong.
\end{proposition}
\begin{longversion}
  \begin{proof}
    If $H$ is an Heyting algebra, then an upset $F \subseteq H$ is a
    filter if it is closed under meets. We recall that filters $F$
    bijectively correspond to congruences $\theta$, by saying $x\theta
    y$ if and only if $x \leftrightarrow y \in F$.

    Thus, if $\f : H \rto H$ is a polynomial, then write it as
    $\eval[]{\phi}(z,\vec{v})$ for a tuple of elements of $\vec{v}$ of
    $H$.

    Let also $H' = H/F$, where $F$ is the upset of $x \impl y$, so
    $\eval[]{\phi}(z,[\vec{v}])$ is a monotone polynomial on $H'$. As
    in $H'$ we have $[x] \leq [y]$, then
    \begin{align*}
      [\f(x)] = [\,\eval[]{\phi}(x,\vec{v})\,] & =
      \eval[]{\phi}([x],[\vec{v}]) \\
      & \leq \eval[]{\phi}([y],[\vec{v}])
      \\
      & = [\,\eval[]{\phi}(y,\vec{v})\,] = [\f(y)].
    \end{align*}
    Yet, the inclusion $[\f(x)] \leq [\f(y)]$ in $H'$ means that
    $\f(x) \impl \f(y)$ is in the principal filter generated by $x
    \impl y$, that is, that equation \eqref{eq:strongenriched} from
    Lemma~\ref{lemma:strong} holds.  \qed
  \end{proof}
\end{longversion}
\begin{proof}
  Recall that the replacement Lemma holds in the \Ipc: $z
  \leftrightarrow w \cons \phi(z) \biimpl \phi(w)$.  Substituting $x$
  for $z$ and $x \land y$ for $w$, and considering that $x \impl y
  \cons x \biimpl (x \land y) $, we derive that $x \impl y \cons
  \phi(x) \biimpl \phi(x \land y)$. Assuming that $u$ is positive in
  $\phi(u)$, we have $ \phi(x) \biimpl \phi(x \land y) \cons \phi(x)
  \impl \phi(x \land y) \cons \phi(x) \impl \phi(y)$, whence $x \impl
  y \cons \phi(x) \impl \phi(y)$.  The last relation immediately
  implies that equation~\eqref{eq:strongenriched} from
  Lemma~\ref{lemma:strong} holds, when $\f$ is a monotone polynomial.
  \qed
\end{proof}
It can be shown that the relation $\f(x) \land y = \f(x\land y)
  \land y$ holds (for any $x,y$ and) for \emph{any} polynomial on a
  \Ha. The analogous remark  for Boolean algebras
  is credited to Peirce, in view of the iteration rule for
  existential graphs of type Alpha, see \cite{Dau2006}.

\begin{proposition}
  \label{prop:fixrimpl}
  If $\f$ is a strong monotone function on $H$ and $a \in H$, then
  \begin{align}
    \label{eq:fixrimpl-conj}
    \mu.a \impl \f & := a \impl \mu.\f\,,
    &
    \mu.a \land \f & := a \land \mu.\f\,.
  \end{align}
\end{proposition}
\begin{proof}
  Let us argue first that first equation holds. To this end, let us
  set $\fea(x) \eqdef a \impl \f(x)$.  From $\f \leq \fea$ we have
  $\Pref_{\fea} \subseteq \Pref_{\f}$. Thus, if $p \in \Pref_{\fea}$,
  then $\mu_{x}.\f(x) = \f(\mu_{x}.\f(x)) \leq \f(p)$ and
  $a \impl\mu.\f \leq a \impl \f(p) = \fea(p) \leq p$. That is,
  $a \impl\mu.\f $ is below any element of $\Pref_{\fea}$. To obtain
  the proposition, we need to argue that $a \impl\mu.\f $ belongs to
  $\Pref_{\fea}$. To this end, we notice that
  $\set{a \impl p \mid p \in \Pref_{\f}} \subseteq \Pref_{\fea}$,
  since if $\f(p) \leq p$, then $\fea(a \impl p) = a \impl \f(a \impl
  p) \leq a \impl \f(p) \leq a \impl p$, where we used the fact that $\f$
is strong,
  thus \eqref{eq:strongimpl} holds.

  \smallskip 

  Let us come now to the second equation, for which we set $\fla(x)
  \eqdef a \land \f(x)$.  Suppose $a \land \f(p) \leq p$, so $\f(p)
  \leq a \impl p$. Then $\f(a \impl p) \leq a \impl \f(p) \leq a \impl
  p$, using \eqref{eq:strongimpl},
  whence $\mu.\f \leq a \impl p$ and $a \land\mu.\f
  \leq p$.
  Thus we are left to argue that $a \land \mu.\f$ is a \pfp of
  $\fla$. Yet, this is true for an arbitrary \pfp $p$ of $\f$: $a
  \land \f(a \land p) \leq a \land \f(p) \leq a \land p$. \qed
\end{proof}

\begin{corollary}
  \label{cor:distrconj}
  For each $n \geq 1$ and each collection $\f_{i}$, $i = 1,\ldots ,n$
  of monotone polynomials, we have the following distribution law:
  \begin{align}
    \mu_{x}. \bigwedge_{i = 1,\ldots ,n} \f_{i}(x) & := \bigwedge_{i =
      1,\ldots ,n} \mu_{x}.\f_{i}(x)\,.
    \label{eq:muconjunction}
  \end{align}
\end{corollary}
\begin{proof}
  For $n = 1$ there is nothing to prove. We suppose therefore that the
  statement holds for every collection of size $n \geq 1$, and prove it holds
  for a collection of size $n+1$. We have
  \begin{align*}
    \mu_{x}.(\f_{n+1}(x) \land \bigwedge_{i=1,\ldots ,n} \f_{i}(x)) &
    := \mu_{x}.\mu_{y}.(\f_{n+1}(y) \land \bigwedge_{i=1,\ldots
      ,n} \f_{i}(x)), \tag*{by \eqref{eq:diag},}\\
      & := \mu_{x}.( (\mu_{y}.\f_{n+1}(y)) \land \bigwedge_{i=1,\ldots
        ,n} \f_{i}(x)), \tag*{by~\eqref{eq:fixrimpl-conj},}\\
      & := (\mu_{y}.\f_{n+1}(y)) \land \mu_{x}.(\bigwedge_{i=1,\ldots
        ,n} \f_{i}(x)), \tag*{again by~\eqref{eq:fixrimpl-conj},}\\
      & := (\mu_{y}.\f_{n+1}(y)) \land \bigwedge_{i=1,\ldots ,n}
      \mu_{x}.\f_{i}(x), \tag*{by the IH. \quad \qed}
  \end{align*}
\end{proof}

The elimination  of \gfp{s} is easy for strong monotone functions
(we are thankful to the
referee for pointing out the following fact, which greatly simplifies our original argument):

\begin{proposition}\label{prop:phitop}
  If $\f : L \rto L$ is any strong monotone function on a bounded
  lattice $L$, then $\f^{2}(\top) = \f(\top)$. Thus $\f(\top)$ is the
  \gfp of $\f$.
\end{proposition}
\begin{proof} Indeed, we have
$\f(\top) = \f(\top) \land \f(\top) \leq \f(\f(\top) \land \top) =
\f^{2}(\top)$.
\qed
\end{proof}

\section{A digression on fixpoints and bisimulation
quantifiers}
\label{sec:quantifiers}
\label{sec:gfp}

The connection between \efp{s} and bisimulation quantifiers, as
emphasized in \cite{DAgostinoHollenberg2000}, was a main motivation to
tackle this research. Although in the end our computations are independant on that, we nevertheless 
want to have a closer look to the topic (the content of this section is not needed afterwards).

It was discovered in \cite{Pitts92} that \Ipc has the uniform
interpolation property. As made clear from the title of
\cite{Pitts92}, this property amounts to an internal existential and
universal quantification. This result was further refined in
\cite{GhilardiZawadowski97} to show that any morphism between finitely
generated \Ha{s} has a left and a right adjoint. We shall be
interested in Heyting algebras $H[x]$ of polynomials with coefficients
from $H$, and to (the left and right adjoints to) the inclusion of $H$
into $H[x]$. The algebra of polynomials $H[x]$ is formally defined as
the coproduct (in the category of \Ha{s}) of $H$ with the free \Ha on
one generator. The universal property gives that if $h_{0} \in H$,
then there exists a unique morphism
$\eval[h_{0}/x]{\cdot} : H[x] \rto H$ such that
$\eval[h_{0}/x]{x} = h_{0}$ and $\eval[h_{0}/x]{h} = h$, for each
$h \in H$. Thus, for $\f \in H[x]$ and $h \in H$, we can define
$\f(h) = \eval[h/x]{\f}$. It follows from \cite{GhilardiZawadowski97}
that if $H$ is finitely generated, then the inclusion
$i_{x} : H \rto H[x]$ has both adjoints
$\exists_{x} ,\forall_{x} : H[x] \rto H$, with
$\exists_{x} \dashv i_{x} \dashv \forall_{x}$. In particular, we shall
use the unit relation for $\exists_{x}$:
\begin{align*}
   \f & \leq i_{x}(\exists_{x}(\f)) \,,\qquad \text{for all $\f \in H[x]$}\,.
\end{align*}
Identifying $h \in H$ with $i_{x}(h) \in H[x]$, we can read the above
inequality as $\f \leq \exists_{x}.\f$.  We can identify a monotone
polynomial, as defined in Definition~\ref{def:monotonepol}, as an
element $\f \in H[x]$ such that $\eval[h_{0}/x]{\f} \leq
\eval[h_{1}/x]{\f}$ whenever $h_{0} \leq h_{1}$.

\begin{proposition}
  If $\f$ is a monotone polynomial on a finitely generated Heyting
  algebra, then
  \begin{align}
    \label{eq:nuexistsone}
    \nu.\f & := \exists_{x}.( x \land (x \impl \f(x)))\,.
  \end{align}
\end{proposition}
\begin{proof}
  By the unit relation $x \land x \impl \f(x) \leq \exists_{x}.(x
  \land x \impl \f(x))$.  Recall that evaluation at $p \in H$ is a
  \Ha morphism, thus it is monotone. Therefore, if $p \in H$ is a
  \pofp of $\f$, then by evaluating the previous inequality at $p$, we
  have
  \begin{align*}
    p & = p\land  p \impl \f(p) \leq  \exists_{x}.(x \land x \impl
    \f(x))\,,
  \end{align*}
  so that $\exists_{x}.(x \land x \impl \f(x))$ is greater than any
  \pofp of $\f$. Let us show that $\exists_{x}.(x \land x
  \impl \f(x))$ is also a \pofp. To this end, it will be
  enough to argue that $x \land x \impl \f(x) \leq \f(\exists_{x}.(x
  \land x \impl \f(x)))$ in $H[x]$. We compute as follows:
  \begin{align*}
    x \land
  x \impl \f(x)
  & \leq \f(x) \land x \impl \f(x) \\
  & \leq \f(x \land x \impl \f(x)), \tag*{since $\f$ is strong, by \eqref{eq:strongconj},} \\
  & \leq \f(\exists_{x}.(x \land x \impl \f(x))), \tag*{since
    $\f$ is monotone. \qed} 
  \end{align*}
\end{proof}

In a similar fashion, we can prove that if $\f$ is a monotone
polynomial on a finitely generated Heyting algebra, then $\mu.\f :=
\forall_{x}.((\f(x) \impl x) \impl x)$.  
As an application, we give an alternative proof of Proposition~\ref{prop:phitop}:

\begin{corollary}
  If $\f$ is a monotone polynomial on a Heyting algebra $H$, then
  \begin{align}
    \label{eq:nuexists}
    \nu.\f & := \f(\top)\,.
  \end{align}
\end{corollary}
\begin{proof}
  It is easy to see that if $\f$ is a monotone polynomial on a
  finitely generated \Ha, then $\exists_{x}.\f = \f(\top)$. Thus we
  have
  \begin{align*}
    \nu.\f & = \exists_{x}.( x \land (x \impl \f(x))) = \exists_{x}.(
    x \land \f(x)) = \top \land \f(\top) = \f(\top)\,. 
  \end{align*}
  Therefore, if $\phi$ is a \fterm whose variables are among set
  $x,y_{1},\ldots ,y_{n}$, then the equation
  $\phi^{2}(\top) = \phi(\top)$ holds in the free \Ha on the set
  $\set{y_{1},\ldots ,y_{n}}$. Consequently, the equation
  $\f(\top) = \f^2(\top)$ holds in $H$, making $\f(\top)$ into the
  \gfp of $\f$.
  \qed
\end{proof}

\section{The elimination procedure}
\label{sec:procedure}

In this Section we present our main result, a procedure that both
axiomatizes and eliminates \lfp{s} of the form $\mu_{x}.\phi(x)$ with
$\phi$ \fpf. Together with the axiomatization of \gfp{s} given in
Section~\ref{sec:gfp}, the procedure can be extended to a procedure to
construct a \fpf formula $\psi$ equivalent to a given formula $\chi$
of the \muIpc.

\begin{definition}
  An occurrence of the variable $x$ is \emph{\stronglypositive} in a
  \fterm $\phi$ if there is no subformula $\psi$ of $\phi$ of the form
  $\psi_{0} \impl \psi_{1}$ such that $x$ is located in $\psi_{0}$. A
  \fterm $\phi$ is \emph{\stronglypositive} in the variable $x$ if
  every occurrence of $x$ is \emph{\stronglypositive} in $\phi$.
  An occurrence of a variable $x$ is \emph{\weaklynegative} in a
  \fterm $\phi$ if it is not \stronglypositive. A \fterm $\phi$ is
  \emph{\weaklynegative} in the variable $x$ if every occurrence of
  $x$ is \emph{\weaklynegative} in $\phi$.
\end{definition}
Observe that a variable might be neither \stronglypositive nor
\weaklynegative in a \fterm. 
A second key concept for the elimination is the following notion of
disjunctive formula.
\begin{definition}
  The set of \fterms that are \emph{disjunctive in the variable $x$}
  is generated by the following grammar:
  \begin{align}
    \phi & \production x \mid \beta \vee \phi \mid \phi \vee \beta
    \mid \alpha \impl \phi \mid \phi \vee \phi\,,
    \label{grammar:disjunctive}
  \end{align}
  where $\alpha$ and $\beta$ are formulas with no occurrence of the
  variable $x$. A \fterm $\phi$ is in \emph{normal form} (w.r.t. $x$)
  if it is a conjunction of \fterms $\phi_{i}$, $i \in I$, so that
  each $\phi_{i}$ either does not contain the variable $x$, or it is
  disjunctive in $x$.
\end{definition}
Notice that disjunctive \fterms are \fullypositive in $x$.  Due to
equation~\eqref{eq:distrimpl} and since the usual distributive laws
hold in \Ha{s}, we have the following Lemma.
\begin{lemma}
  Every \spositive \fterm is equivalent to a \fterm in normal form.
\end{lemma}

In order to compute the \lfp $\mu_{x}.\phi$, we 
take the following steps:
\begin{enumerate}
\item We rename all the \wnegative occurrences of $x$ in $\phi$ to a
  fresh variable $y$, so $\phi(x) = \psi(x,x/y)$ with $\psi$
  \stronglypositive in $x$ and \weaklynegative in $y$.
\item We compute a normal form of $\psi(x,y)$, so this formula is
  equivalent to a conjunction $\bigwedge_{i \in I} \psi_{i}(x,y)$ with
  each $\psi_{i}$ disjunctive in $x$ or not containing the variable
  $x$.
\item \emph{\Spositive elimination.} For each $i \in I$:
  if $x$ has an occurrence in $\psi_{i}$, we compute then a formula
  $\psi'_{i}$ equivalent to the \lfp $\mu_{x}.\psi_{i}(x,y)$ and
  observe that $\psi'_{i}$ is \weaklynegative in $y$; otherwise, we
  let $\psi'_{i} = \psi_{i}$.
  \item \emph{\Wnegative elimination}. The formula $\bigwedge_{i \in
      I} \psi'_{i}(y)$ is \wnegative in $y$; we compute a formula
    $\chi$ equivalent to $\mu_{y}.\bigwedge_{i} \psi'_{i}(y)$ and
    return it.
\end{enumerate}
The correction of the procedure relies on the following chain of
equivalences:
\begin{align*}
  \mu_{x}.\phi(x) & = \mu_{y}.\mu_{x}.\psi(x,y) 
  = \mu_{y}.\mu_{x}.\bigwedge_{i \in I}\psi_{i}(x,y), \tag*{where we use \diag,}\\
  & =
  \mu_{y}.\bigwedge_{i \in I}\mu_{x}.\psi_{i}(x,y) 
  = \mu_{y}.\bigwedge_{i \in I}\psi'_{i}(y) = \chi,
  \tag*{where we have used \eqref{eq:muconjunction}.}
\end{align*}

\subsection{\Spositive elimination}

We tackle here the problem of computing the \lfp $\mu_{x}.\phi$ of a
\fterm $\phi$ which is \emph{disjunctive} in $x$.
Recall that the formulas $\alpha$ and $\beta$ appearing in a parse
tree as leaves---according to the
grammar~\eqref{grammar:disjunctive}---do not contain the variable $x$.
We call such a formula $\alpha$ a \emph{head subformula} of $\phi$,
and such a $\beta$ a \emph{side subformula} of $\phi$, and thus we
put:
\begin{align*}
  \Head(\phi) & \eqdef \set{ \alpha \mid \alpha \text{ is a head
      subformula of } \phi } \,,\\
  \Side(\phi) & \eqdef \set{ \beta \mid \beta \text{ is a side
      subformula of } \phi } \,.
\end{align*}

Recall that a monotone function $f : P \rto P$ is \emph{inflating} if $x \leq f(x)$. 
\begin{lemma}
  The interpretation of a \stronglypositive disjunctive formula $\phi$
  as a function of $x$ is inflating.
\end{lemma}
The key observation needed to prove Proposition~\ref{prop:mudformula}
is the following Lemma on monotone inflating functions. In the
statement of the Lemma we assume that $P$ is a join-semilattice, and
that $f \vee g$ is the pointwise join of the two functions $f$ and
$g$.
\begin{lemma}
  \label{lemma:circvee}
  If $f,g:P \rto P$ are monotone inflating functions, then $\Pref_{f
    \vee g} = \Pref_{f \circ g}$.
  Consequently,   for any monotone function $h : P \rto P$,
  we have
  \begin{align}
    \label{eq:circvee}
    \mu.(\,f \vee g \vee h\,) & :=: \mu.(\,(f \circ g) \vee h\,)\,.
  \end{align}
\end{lemma}
\begin{proof}
  Observe firstly that $\Pref_{f \vee g}= \Pref_{f} \cap
  \Pref_{g}$.
  If $p \in \Pref_{f \circ g}$, then $f(p) \leq f(g(p)) \leq p$ and $g(p) \leq f(g(p)) \leq p$,
  showing that $p \in \Pref_{f \vee g}$.
  Conversely, if $p \in \Pref_{f \vee g}$, then $p$ is a fixed point
  of both $f$ and $g$, since these functions are inflating. It follows
  that $f(g(p)) = f(p) = p$,
  showing $p \in \Pref_{f \circ g}$. 
  
  We have argued that $\Pref_{f \vee g}$ coincides with
  $\Pref_{f \circ g}$; this implies that
  $\Pref_{(f \circ g) \vee h} = \Pref_{f \vee g \vee h}$ and, from
  this equality, equation~\eqref{eq:circvee} immediately follows.
  \qed
\end{proof}

To ease reading of the next Proposition and of its proof, let us put
\begin{align*}
  \Nec[\alpha] \phi & \eqdef \alpha \impl \phi\,.
\end{align*}
\begin{proposition}
  \label{prop:mudformula}
  If $\phi$ is a disjunctive \fterm, then
  \begin{align}
      \label{eq:mudformula}
      \mu.\phi & = 
    \Nec[\bigwedge_{\alpha \in \Head(\phi)} \alpha](\bigvee_{\beta \in
    \Side(\phi)} \beta)\,.
  \end{align}
\end{proposition}
\begin{proof}
  For $\psi,\chi$ \fterms, let us write $\psi \sim \chi$ when $\mu.\psi =
  \mu.\chi$.
  We say that a disjunctive formula $\psi$ is reduced (w.r.t. $\phi$)
  if either it is $x$, or it is of the form $\beta \vee x$ (or $x \vee
  \beta$) for some $\beta \in \Side(\phi)$, or of the form
  $\Nec[\alpha]x$ for some $\alpha \in \Head(\phi)$.
  A set $\Phi$ of disjunctive formulas  is \emph{reduced} if every formula in
  $\Phi$ is reduced.
  
  We shall compute a reduced set of disjunctive formulas $\Phi_{k}$
  such that $\phi \sim \bigvee \Phi_{k}$.
  Thus let $\Phi_{0} = \set{\phi}$.  If $\Phi_{i}$ is not reduced,
  then there is $\phi_{0} \in \Phi_{i}$ which is not reduced, thus of
  the form (a) $\beta \vee \psi$ (or $\psi \vee \beta$) with $\psi
  \neq x$, or (b) $\Nec[\alpha]\psi$ with $\psi \neq x$, or (c)
  $\psi_{1} \vee \psi_{2}$.  According to the case ($\ell$), with $\ell \in
  \set{a,b,c}$, we let $\Phi_{i + 1}$ be $(\Phi_{i} \setminus
  \set{\phi_{0}}) \cup \Psi_{\ell}$ where $\Psi_{\ell}$ is as follows:
  \begin{align*}
    \Psi_{a} &= \set{\beta \vee x, \psi}, 
    & \Psi_{b} & = \set{\Nec[\alpha]x, \psi}, 
    & \Psi_{c} & = \set{\psi_{1},\psi_{2}}\,.
  \end{align*}
  By Lemma~\ref{lemma:circvee}, we have $\bigvee \Phi_{i} \sim \bigvee
  \Phi_{i+1}$. Morever, for some $k \geq 0$, $\Phi_{k}$ is reduced and
  $\Phi_{k} \subseteq \set{\Nec[\alpha]x \mid \alpha \in \Head(\phi)}
  \cup \set{\beta \vee x \mid \beta \in \Side(\phi)} \cup
  \set{x}$. Consequently
  \begin{align}
    \label{eq:allsideQM}
    \mu_{x}.\phi(x) & = \mu_{x}.\bigvee \Phi_{k} \leq \mu_{x}.(x \vee
    \bigvee_{\alpha \in \Head(\phi)} \Nec[\alpha]x \vee \bigvee_{\beta
      \in \Side(\phi)} \beta \vee x )\,.
  \end{align}
  On the other hand, if $\alpha \in \Head(\phi)$, then $\phi(x) =
  \psi_{1}(x,\Nec[\alpha]\psi_{2}(x))$ for some disjunctive formulas
  $\psi_{1}$ and $\psi_{2}$, so
  \begin{align*}
    \Nec[\alpha]x & \leq \Nec[\alpha]\psi_{2}(x) \leq
    \psi_{1}(x,\Nec[\alpha]\psi_{2}(x)) = \phi(x)
  \end{align*}
  and, similarly, $\beta \vee x  \leq \phi(x)$, whenever $\beta \in \Side(\phi)$.
  It follows that
  \begin{align*}
    x \vee \bigvee_{\alpha \in \Head(\phi)} \Nec[\alpha]x \vee
    \bigvee_{\beta \in \Side(\phi)} \beta \vee x & \leq \phi(x)\,,
  \end{align*}
  whence, by taking the \lfp in both sides of the above inequality, we
  derive equality in \eqref{eq:allsideQM}.
  Finally, in order to obtain \eqref{eq:mudformula}, we compute as follows:
  \begin{align*}
    \smalllhs[7mm]{\mu_{x}.(x \vee \bigvee_{\alpha \in \Head(\phi)}
      \Nec[\alpha]x \vee
      \bigvee_{\beta \in \Side(\phi)} \beta \vee x)} \\
    & = 
    \mu_{x}.(\Nec[\alpha_{1}] \ldots \Nec[\alpha_{n}]x \vee
    (x \vee \bigvee_{\beta \in \Side(\phi)} \beta \vee x)) 
    \tag*{
by Lemma~\ref{lemma:circvee}, with  $\Head(\phi) = \set{\alpha_{1},\ldots ,\alpha_{n}}$,}
    \\
    & = \mu_{x}.(\Nec[\bigwedge_{\alpha \in \Head(\phi)} \alpha]x \vee
    (x \vee \bigvee_{\beta \in \Side(\phi)} \beta \vee x)), 
    \tag*{
      since $\Nec[\alpha_{1}] \ldots \Nec[\alpha_{n}]x =
        \Nec[\bigwedge_{i = 1,\ldots ,n}\alpha_{i}]x$,}
    \\
    & = \mu_{x}.(\Nec[\bigwedge_{\alpha \in \Head(\phi)} \alpha]( x
    \vee \bigvee_{\beta \in \Side(\phi)} \beta \vee x)) , \tag*{by
      Lemma~\ref{lemma:circvee},}
    \\
    & = \Nec[\bigwedge_{\alpha \in \Head(\phi)} \alpha]\mu_{x}.( x
    \vee \bigvee_{\beta \in \Side(\phi)} \beta \vee x) , \tag*{by
      Proposition~\ref{prop:fixrimpl},}
    \\
    & = \Nec[\bigwedge_{\alpha \in \Head(\phi)} \alpha](
    \bigvee_{\beta \in \Side(\phi)} \beta)\,.  \tag*{\qedhere}
  \end{align*}
\end{proof}

\subsection{\Wnegative elimination}
\label{sec:weaklynegative}

If $\phi$ is \wnegative in $x$ 
then we can write
\begin{align}
  \label{eq:wndecomposition}
  \phi(x) & = \psi_{0}(\psi_{1}(x), \ldots ,\psi_{n}(x))\,,
\end{align}
for \fterms $\psi_{0}(y_{1},\ldots ,y_{n})$ and $\psi_{i}(x)$, $i =
1,\ldots ,n$, such that: (a) all the variables $y_{i}$ are negative in $\psi_{0}$; (b) for $i = 1,\ldots ,n$,
$x$ is negative  $\psi_{i}$.

\begin{proposition}
  Let $\langle \nu_{1},\ldots ,\nu_{n}\rangle$ be a collection of
  \fterms denoting the greatest solution of the system of equations
  $\set{ y_{i} = \psi_{i}(\psi_{0}(y_{1},\ldots ,y_{n})) \mid i =
    1,\ldots ,n }$. Then $\psi_{0}(\nu_{1},\ldots ,\nu_{n})$ is a
  formula equivalent to $\mu_{x}.\phi(x)$.
\end{proposition}
\begin{proof}
  Let $v : \Vars \setminus \set{x,y_{1},\ldots ,y_{n}} \rto H$ be a
  partial valuation into an \Ha $H$, put $\f_{0} = \eval{\psi_{0}}$
  and, for $i = 1,\ldots ,n$, $\f_{i} = \eval{\psi_{i}}$. Then
  $\f_{0}$ is a monotone function from $ [ H^{op}]^{n}$ to $H$. Here
  $H^{op}$ is the poset with the same elements as $H$ but with the
  opposite ordering relation. Similarly, for $1 \leq i \leq n$,
  $\f_{i} : H \rto H^{op}$.  If we let $\bar{\f} = \langle \f_{i} \mid
  i = 1,\ldots ,n \rangle \circ \f_{0}$, then $\bar{\f}
  :{[H^{op}]}^{n} \rto {[H^{op}]}^{n}$.
  We exploit next the fact that $(\cdot)^{op}$ is a functor, so that
  $f^{op} : P^{op} \rto Q^{op}$ is the same monotone function as $f$,
  but considered as having distinct domain and codomain.  Then, using
  \roll, we can write
  \begin{align}
    \notag
    \label{eq:munu}
    \mu. (\,\f_{0} \circ \langle \f_{i} \mid i = 1,\ldots ,n \rangle
    \,) & = \f_{0}(\, \langle \f_{i} \mid i = 1,\ldots ,n \rangle
    \circ \f_{0}\,)
    \\
    & = \f_{0}(\,\mu.\bar{\f}\,) = \f_{0}(\,\nu.\bar{f}^{op}\,)\,,
  \end{align}
  since the \lfp of $f$ in $P^{op}$ is the \gfp of $f^{op}$ in
  $P$. That is, if we consider the function $\langle \f_{i} \mid i =
  1,\ldots ,n \rangle \circ \f_{0}$ as sending a tuple of elements of
  $H$ (as opposite to $H^{op}$) to another such a tuple, then
  equation~\eqref{eq:munu} proves that a formula denoting the \lfp of
  $\phi$ is constructible out of formulas for the greatest solution of
  the system mentioned in the statement of the Proposition.  \qed
\end{proof}

As far as computing the greatest solution of the system mentioned in
the Proposition, this can be achieved by using the \Bekic elimination
principle, see Lemma~\ref{lemma:bekic}.  This principle implies that
solutions of systems can be constructed from solutions of linear
systems, i.e. from usual parametrized \fp{s}. In our case, as
witnessed by equation~\eqref{eq:nuexists}, these parametrized \gfp{s}
are computed by substituting $\top$ for the \fp variable. In the next
Section we shall give a more explicit description, by means of
approximants, of the \lfp of a \weaklynegative formula $\phi$.

\section{Upper bounds on closure ordinals}
\label{sec:upperbounds}

Recall that Ruitenburg's result \cite{Ruitenburg84} implies that a
monotone formula converges to its (parametrized) \lfp by iterating the
formula $n$ times from $\bot$, for some $n \geq 0$. That is, we can
always substitute $\mu_{x}.\phi(x)$ for some equivalent
$\phi^{n}(\bot)$. We show, in this Section, how to extract, from the
procedure just seen, upper bounds for such a number $n$.
\begin{proposition}
  \label{prop:convergencedf}
  If $\phi$ is a disjunctive formula and $n$ is the cardinality of the
  set $\Head(\phi)$, then 
  \begin{align}
    \label{eq:convergencedf}
    \mu_{x}.\phi(x)  & = \phi^{n +1}(\bot)\,.
  \end{align}
\end{proposition}
\begin{proof}
  We have seen, in the proof of Propositon~\ref{prop:mudformula}, that
  $\Nec[\alpha]x \leq \phi(x)$ for any $\alpha \in \Head(\phi)$ and,
  similarly, $\beta \vee x \leq \phi(x)$ for any $\beta \in
  \Side(\phi)$.  Thus we have
  \begin{align*}
    \bigvee_{\beta \in \Side(\phi)} \beta
    & = \bigvee_{\beta \in \Side(\phi)} \beta \vee \bot \leq \phi(\bot)\,.
  \end{align*}
  Let $\Head(\phi) = \set{\alpha_{1},\ldots ,\alpha_{n}}$. Supposing
  that
  $\Nec[\alpha_{i}] \ldots \Nec[\alpha_{1}](\bigvee_{\beta \in
    \Side(\phi)} \beta) \leq \phi^{i +1}(\bot)$,
  then
  \begin{align*}
   \Nec[\alpha_{i+1}] \Nec[\alpha_{i}] \ldots \Nec[\alpha_{1}](\bigvee_{\beta \in
      \Side(\phi)} \beta) \leq \Nec[\alpha_{i+1}](\phi^{i +1}(\bot))
    \leq\phi(\phi^{i +1}(\bot)) = \phi^{i +2}(\bot)\,.
  \end{align*}
  Whence
  \begin{align*}
    \mu_{x}.\phi(x) &
    = \Nec[\bigwedge_{i = 1,\ldots, n} \alpha_{i}](\bigvee_{\beta \in
      \Side(\phi)} \beta) 
    = \Nec[\alpha_{n}]  \ldots \Nec[\alpha_{1}](\bigvee_{\beta \in
      \Side(\phi)} \beta) \leq \phi^{n + 1}(\bot)\,.
    \tag*{\qedhere}
  \end{align*}
\end{proof}
The upper bound given in 
\eqref{eq:convergencedf} is optimal: if we let $\phi_{n}(x) \eqdef b
\vee \bigvee_{i=1,\ldots ,n} a_{i} \impl x$ and consider the \Ha of
downsets of $\langle P(\set{1,\ldots ,n}),\subseteq \rangle$, then,
interpreting $b$ as $\set{\emptyset}$ and $a_{i}$ as $\set{ s
  \subseteq \set{1,\ldots ,n} \mid i \not\in s}$, $\phi_{n}$ converges
exactly after $n + 1$ steps.

\medskip

In order to tackle convergence of \weaklynegative formulas, we mention
some general statements, where we assume that all the posets have a
least element.
\begin{lemma}\textrm{Convergence for \roll.}
  \label{lemma:convrolling}
  Let $f : P \rto Q$ and $g: Q \rto P$ be monotone functions. If
  $\mu. (f \circ g) = (f \circ g)^{n}(\bot)$, then $\mu. (g \circ f) =
  (g \circ f)^{n+1}(\bot)$.
\end{lemma}

\begin{lemma}\label{lemma:convdiag}\textrm{Convergence for \diag.}
  Let $f : P \times P \rto P$ be a monotone function.
  For each $p \in P$, put $g_{p}(y) = f(p,x)$ and $h(x) =
  \mu_{y}.g_{x}(y)$.  Suppose that, for each $p \in P$, $h(p) =
  \mu_{y}.f(p,y) =g_{p}^{n} (\bot)$ and that $\mu_{x}.h(x) =
  h^{m}(\bot)$.  Then $\mu_{x}.f(x,x) = f^{nm}(\bot,\bot)$.
\end{lemma}

For our purposes, the following Lemma provides more accurate upper
bounds than Lemma~\ref{lemma:convdiag}.
\begin{lemma}
  Let $f,g : H \rto H$ be strong monotone mappings. If
  $\mu.f = f^{n}(\bot)$ and $\mu.g = g^{m}(\bot)$, then
  $\mu.f \land g = (f \land g)^{n +m -1}(\bot)$.
\end{lemma}

For the \Bekic property we have a similar statement, bounding
convergence of $\langle f,g\rangle$ by $(n + 1)(m+1) -1$, with $m$ and
$n$ being bounds on convergence of $\mu_{y}.g(x,y)$ and
$\mu_{x}.f(x,\mu_{y}.g(x,y))$, respectively. While in general this
bound is optimal, the relevant observation is, for our purposes, the
following Lemma.
\begin{lemma}
  \label{lemma:convergencesystem}
  Let $\{ x_{i} = f_{i}(x_{1},\ldots ,x_{k}) \mid i = 1,\ldots ,k \}$
  be a monotone system of equations $P$ on some poset with least
  element $\bot$. Suppose that all the functions generated under
  substitution from $\set{ f_{1},\ldots ,f_{k} } \cup \set{\bot}$
  converge to their parametrized \lfp in one step. Then the \lsol of
  this system of equations is obtained by iterating $k$ times $\langle
  f_{1},\ldots ,f_{k} \rangle$ from $(\bot,\ldots ,\bot) \in P^{k}$.
\end{lemma}

\begin{proposition}
  \label{prop:convergesforwefs}
  Let $\phi(x)$ be a weakly negative formula, so that we have a
  decomposition of the form \eqref{eq:wndecomposition}.  Then
  $\phi(x)$ converges at its \lfp in at most $n + 1$ steps.
\end{proposition}
\begin{proof}
  Applying Lemma~\ref{lemma:convergencesystem}, we have 
  \begin{align}
    \nu.(\langle \psi_{i} \mid
    i = 1,\ldots ,n \rangle \circ \psi_{0}) & = (\langle \psi_{i} \mid
   i = 1,\ldots ,n \rangle \circ \psi_{0})^{n}(\top)\,.
   \label{eq:munu2}
 \end{align}
 Considering that
 \begin{align*}
   \mu.\phi & = \mu.(\psi_{0} \circ \langle \psi_{i} \mid
   i = 1,\ldots ,n \rangle) 
   = \psi_{0} (\nu.(\langle \psi_{i} \mid
      i = 1,\ldots ,n \rangle \circ \psi_{0}))
   \end{align*}
   we can use  \eqref{eq:munu2} and
   Lemma~\ref{lemma:convrolling} to deduce that
   \begin{align*}
     \mu.\phi & = (\psi_{0} \circ \langle \psi_{i} \mid
     i = 1,\ldots ,n \rangle)^{n+1}(\bot) \,.
     \tag*{\qedhere}
   \end{align*}
\end{proof}

It is possible to combine Propositions~\ref{prop:convergencedf} and
\ref{prop:convergesforwefs} with Lemma~\ref{lemma:convrolling} to
obtain upper bounds for all formulas. Yet, mainly due to the
exponential blow-up in computing an equivalent normal-form of a given
formula, that is, step 2 of the procedure described in
Section~\ref{sec:procedure}, these bounds turn out to be exponential functions of the size of the formula. It is possible on
the other hand to pinpoint fragments of the \muIpc for which we still
have polynomial bounds. For example, if we define a \fterm to be
\wdisjunctive if it is generated by the grammar
\eqref{grammar:disjunctive}, with the difference that we allow $x$ to
have \wnegative occurrences in $\alpha$ and $\beta$, then bounds are
polynomials of order $2$.

\section{Conclusions}
\label{sec:conclusions}

As mentioned in the Introduction, a main motivation for the research
described in this paper was 
to provide in-depth answers to
the question of why
alternation-depth hierarchies in $\mu$-calculi collapse or are
trivial. Until now, the authors
dealt with 
trivial alternation-depth hierarchies. The tools and
ideas so far developed still need to be tested when a hierarchy does
not completely collapse at its lowest level. In particular, and given
the closeness of \IL with Modal Logic based on transitive frames, it
becomes appealing to investigate further connections with existing
work on the subject
\cite{AlberucciFacchini09,AlberucciFacchini09b,DAgostinoLenzi2010,Visser05}.

Compared to other works, such as \cite{Mardaev1993,Mardaev1994}, we
definitely took an algebraic and constructive approach to the problem
of showing definability of \lfp{s} within the \Ipc. 
Witnessing the fruitfulness of our approach, the algebra made the goal
of computing upper bounds of closure ordinals of the monotone
functions denoted by intuitionisitc formulas an accessible task.
\blue{Let us notice on the way that our work leads to an obvious
  \emph{decision procedure}, based on any decision procedure for \Ipc,
  for the \muIPC. This logic, already studied on the side of proof
  theory and of game semantics \cite{Clairambault13}, should also be
  of interest in verification, for example when transition systems
  come with some ordering and upward or downward closed properties are
  defined by $\mu$-formulas, see \cite{BS2013}.}

Overall, we believe that understanding \efp{s} and more in general
\fp{s} in an intuitionisitc setting---where sparse but surprising
results are known, see for \cite{BauerLumsdaine13} example---is still
in quest for an elementary but solid theory to be developed. The
present paper is a contribution toward this goal.

\bibliographystyle{splncs03}
\bibliography{biblio}

\end{document}